\documentclass[11pt]{article} 
\usepackage{amsfonts,amscd,amsmath,graphicx}

\def\NP#1#2{ Nucl.Phys. B#1 (#2)} 
\def\PL#1#2{ Phys.Lett. B#1 (#2)}

\def\PR#1#2{Phys.Rev. D#1 (#2)} 
\def\IJMP#1#2{ Int.J.Mod.Phys. A#1 (#2)}

\def\HP#1#2{ JHEP #1 (#2)} 
\def\ap{ \alpha^{\prime}} 
\def\ai{\pi\alpha^{\prime}}
\def\pd{\partial}

\def\wp{w^{\prime}}
\def\yp{y^{\prime}}
\def\pnu{\nu^{\prime}}
\def\pr{\frak G}
\newcommand{\zm}{{}_\perp}
\newcommand{\ep}{\text e}
\newcommand{\oh}{\frac{1}{2}}

\def\3{\Phi^3}
\def\w{\text{w}}
\def\pw{\text{w}^\prime}
\def\g{\text g}

\title{Diagrams of Noncommutative $\3$ Theory from String Theory}
\author{Oleg Andreev\thanks{e-mail:  andreev@physik.hu-berlin.de}
\thanks{Permanent address: Landau Institute, Moscow, Russia}
\hspace{.15cm} and Harald Dorn\thanks{e-mail: dorn@physik.hu-berlin.de}
\\ \\
Humboldt--Universit\"at zu Berlin, Institut f\"ur Physik\\
Invalidenstra\ss e 110, D-10115 Berlin, Germany}

\date{}
\begin{document} 

\maketitle 
\begin{abstract} 
Starting from tree and one-loop tachyon amplitudes of open string theory in the presence of a 
constant $B$-field, we explore two problems. First we show that in the noncommutative field theory limit 
the amplitudes reduce to tree and one-loop diagrams of the noncommutative $\3$ theory. Next, we check 
factorization of the one-loop amplitudes in the long cylinder limit.
\\
PACS : 11.25.-w, 11.25.Db   \\
Keywords: strings, noncommutative field theory
\end{abstract}

\vspace{-10cm}
\begin{flushright}
hep-th/0003113      \\
HU Berlin-EP-00/19
\end{flushright}
\vspace{9cm}


\section{ Introduction} 
\renewcommand{\theequation}{1.\arabic{equation}}
\setcounter{equation}{0}

The fact that string theory can not only be used as a candidate for unified theories but as a powerful tool 
for computing perturbative field theory amplitudes is known since the early seventies. In the simplest case of 
a scalar field Scherk showed how to derive tree and one-loop diagrams of the $\3$ theory from the dual 
model (pre-string theory) \cite{Scherk, dual-rev}. Later this approach was extensively used in the 
framework of string theory by many people (see, e.g., [3-8], and \cite{DiV-rev} for a review). Recently, the 
idea that the spacetime coordinates do not commute draw much attention (see \cite{SW} and a list of 
references therein) . On the one hand, scalar noncommutative field theories were studied in [11-18]. On the 
other hand, it was realized that 
noncommutative geometry naturally appears in the framework of open string theory in the presence of 
a constant $B$-field. The purpose of this paper is to show that tree and one-loop diagrams of the 
noncommutative $\3$  theory can be also derived from string amplitudes.

Before starting our discussion of one loop diagrams of the noncommutative $\3$ theory, we will make a 
detour and discuss open strings in the presence of a constant $B$-field at the tree level (see, e.g., \cite{SW} and 
references therein).

In this case the world-sheet action is given by 
\begin{equation}\label{ac0}
S=\frac{1}{4\ai }\int_{D}d^2z\,\bigl(
g_{ij}\pd_aX^i\pd^aX^j-2i\ai B_{ij}\varepsilon^{ab}\pd_aX^i\pd_bX^j\bigr)\,\,+\,\,\varphi
\quad,
\end{equation}
where $D$ means the string world-sheet, namely a disk. $g_{ij}$, $B_{ij}$, $\varphi$ are the 
constant metric, antisymmetric tensor and dilaton fields, respectively. $X^i$ map 
the world-sheet to the target space ($Dp$-brane) and $i,\,j=1,\dots , d=p+1$. The world-sheet indices are 
denoted by $a,b$. The disk can be, of course, conformally mapped to the upper half plane i.e., the region 
$\text{Im}\,\w\geq 0$ on the complex plane whose coordinate is $\w$. 

To analyze open string theory defined by the  world-sheet action \eqref{ac0}, one first has to determine the 
propagator. To do so, it is necessary to define the boundary conditions. They are \footnote{For the sake 
of simplicity, we use the matrix notations here and below.}
\begin{equation}\label{bc0}
g(\pd -\bar\pd )\pr (\w, \pw)+2\ai  B(\pd+\bar\pd)\pr(\w,\pw )=0 \quad\text{for}\quad\text{Im}\,\w=0
\quad.
\end{equation}
Here $\pr ^{ij}(\w,\pw )=\langle X^i(\w)X^j(\pw)\rangle$ and $\pd=\pd/\pd\w,\,
\bar\pd=\pd/\pd\bar{\w}$.

Evaluated at boundary points, the propagator with these boundary conditions is \cite{AC,SW} 
\begin{equation}\label{pr0}
\pr(s,s^{\prime})=-2\ap G^{-1}\ln\left\vert s-s^{\prime}\right\vert+
\frac{i}{2}\theta\epsilon(s-s^{\prime})\quad,
\end{equation}
where 
\begin{equation}\label{nv}
G=(g-2\ai B)g^{-1}(g+2\ai B)
\quad,\quad
\theta=-(2\ai )^2(g+2\ai B)^{-1}B(g-2\ai B)^{-1}
\quad.
\end{equation}
 $s=\text{Re}\,\w$ and $\epsilon(s)$ is the step function that is $1$ or $-1$ for positive or 
negative $s$. Following \cite{SW}, we will refer to $g,B$ as closed string parameters (variables) 
and $G, \theta$ as open string parameters. There is an interesting point that we should mention 
about the tree level. The propagator between boundary points depends only on the open string parameters. 

Let us now define the open string tachyonic vertex operator \footnote{We will give some motivations 
for such a definition in the next section.} 
\begin{equation}\label{tach}
V(k)=\int ds\,\ep^{ik\cdot X}
\quad,
\end{equation}
where $A\cdot B\equiv 
A_iB^i$. A simple analysis shows that the vertex operator (its integrand) is a primary field of 
conformal dimension one as long as $\ap kG^{-1}k=1$.
 
For M open string tachyons, the tree amplitude is given by 
\begin{equation}\label{amp0}
\begin{split}
A_M=A(k_1,\dots ,k_M)&={\cal N}_0 \left(\ap\right)^{\Delta} G_s^M\,
Tr(\lambda_1\dots\lambda_M)\,{\cal V}_d
\langle V(k_1)\dots V(k_M)\rangle 
\\
&\phantom{=}+\text{noncyclic permutations}
\quad, \quad
\text{where}\quad \langle \,\dots\,\rangle=\int {\cal D}X^{\prime}\,e^{-S}\,\dots\quad . 
\end{split}
\end{equation}
Here $\Delta=\frac{d-2}{4}M-\frac{d}{2}$. We split the integral into the integral over the zero 
mode $\text{\footnotesize  X}$ and the integral 
over nonzero modes. We use the following measure for the integral over the zero mode \footnote{The 
zero mode integration includes the $\text{\footnotesize  X}$-dependence in $\langle\,\,\,\rangle$ and gives 
a factor $\delta\left(\sum k_i\right)$.}
\begin{equation}\label{zm}
{\cal V}_d=\int d^{\,p+1}\text{\footnotesize  X}\sqrt G
\quad.
\end{equation}
This assumes that the dilaton field is redefined as \cite{SW}
\begin{equation}\label{dilaton}
\hat\varphi =\varphi+\oh\ln\det\left(G\left(g+2\ai B\right)^{-1}\right)
\quad.
\end{equation}
However, we have defined the open string coupling as $G_s^2=\ep^{\hat\varphi}$ rather than 
$G_s=\ep^{\hat\varphi}$ as it was done in \cite{SW}.

Moreover, by dividing the invariant measure of the M\"obius group in \eqref{amp0}, three vertex 
operators are fixed to arbitrary positions on the boundary. Each vertex operator is related to a factor $G_s$ 
together with the Chan-Paton degrees of freedom. Thus the amplitude has the appropriate trace factor. 
${\cal N}_0$ is a normalization constant (see, e.g., \cite{DiV}).

In fact, there are two possibilities in taking the limit $\ap\rightarrow 0$ \cite{SW}. The first is to do so while 
keeping the closed string parameters $g,B$ fixed. As a result, in this case one expects the ordinary 
$\3$ theory. This is the standard field theory limit. The second one is to keep 
the open string theory parameters $G,\theta$ fixed \footnote{Both limits assume that 
the corresponding tachyon mass is treated as a free, but fixed parameter.}. The expected result now is 
the noncommutative 
$\3$ theory. This is the noncommutative field theory limit or Seiberg-Witten limit. Since we are 
interested in the noncommutative field theory we will mainly discuss the noncommutative field theory limit. 
 
The tree-tachyon amplitudes already show that the noncommutative field theory limit of string 
amplitudes corresponds to tree-diagrams of the noncommutative $\3$ theory with colour indices
\begin{equation}\label{3}
\hat S=Tr\int d^{p+1}\text{\footnotesize  X}\sqrt{\det G}\left(\oh\pd_i\Phi\pd^i\Phi+\oh m^2\Phi^2+
\frac{1}{6}\g\,\Phi\ast\Phi\ast\Phi\right)
\quad,
\end{equation}
where the coupling constant $\g$ depends on the open string coupling and 
string parameter as $G_s\left(\ap\right)^{(d-6)/4}$. The $\ast$-product is defined by 
\begin{equation}\label{star}
f(\text{\footnotesize  X})\ast\psi (\text{\footnotesize  X})=\ep^{\frac{i}{2}\theta^{ij}\frac{\pd}{\pd y^i}
\frac{\pd}{\pd z^j}}f(\text{\footnotesize  X}+y)\psi (\text{\footnotesize  X}+z)
\quad.
\end{equation}

\section{Open string one-loop amplitudes in the presence of constant $B$-field } 
\renewcommand{\theequation}{2.\arabic{equation}}
\setcounter{equation}{0}

\subsection{General analysis}
The world-sheet action is now given by 
\begin{equation}\label{ac}
S=\frac{1}{4\ai }\int_{C_2}d^2w\,\bigl(
g_{ij}\pd_aX^i\pd^aX^j-2i\ai B_{ij}\varepsilon^{ab}\pd_aX^i\pd_bX^j\bigr)
\quad,
\end{equation}
where $C_2$ denotes the string world-sheet for the one-loop orientable open string i.e., a cylinder (annulus). 
Note that there is no a constant dilaton field as the Euler characteristic of the cylinder is zero.
We describe $C_2$ as the region 
\begin{equation*}
0\leq \text{Re}\, w\,\leq 1
\quad,\quad
w\equiv w+2i\tau
\end{equation*}
on the complex plane whose metric is $ds^2=dwd\bar w$. A flat annulus with inner 
radius $a$ and outer radius $b$ can be obtained from the cylinder by 
\begin{equation*}
z=a\exp{(-w\ln q )}\quad,
\end{equation*}
where the 
modular parameter of the annulus is given by  $q=a/b=\exp{(-\frac{\pi}{\tau})}$. 

To analyze open string theory defined by the  world-sheet action \eqref{ac}, we take a slight modification 
of the propagator found in \cite{AC}. So, we define the boundary conditions as 
\begin{equation}\label{bc}
g\frac{\pd}{\pd x}\pr (w,\wp )-2\ai iB\frac{\pd}{\pd y}\pr(w,\wp )=
\begin{cases}
\,\,\,\,\frac{\ai }{2\tau}\quad\text{for}\quad x=0\,,\\
-\frac{\ai }{2\tau}\quad\text{for}\quad x=1\,.
\end{cases}
\end{equation}
Here $w=x+iy$. The propagator with these boundary conditions is then 
\begin{equation}\label{pr}
\begin{split}
\pr(w,\wp )&=
-\ap g^{-1}\ln\left\vert q^{\oh(\wp -w)}-q^{\oh(w-\wp)}\right\vert-
2\ap G^{-1}\sum_{n=1}^{\infty}\ln
\left[\,
\left\lvert 1-q^{2n-2+w+\bar\wp}\right\rvert
\left\lvert 1-q^{2n-w-\bar w^{\prime}}\right\rvert^{\phantom{\vert}}
\right]
\\
&\phantom{=}-
\ap g^{-1}\sum_{n=1}^{\infty}\ln
\left[
\left\lvert 1-q^{2n+\wp-w}\right\rvert
\left\lvert 1-q^{2n+w-\wp}\right\rvert
\left\lvert 1-q^{2n-2+w+\bar\wp}\right\rvert^{-1}
\left\lvert 1-q^{2n-w-\bar\wp}\right\rvert^{-1}
\right]
\\
&\phantom{=}-\frac{1}{2\pi}\theta\sum_{n=1}^{\infty}\ln\left[
\left(1-q^{2n-2+w+\bar\wp}\right)
\left(1-q^{2n-w-\bar\wp}\right)
\left(1-q^{2n-2+\bar w+\wp}\right)^{-1}
\left(1-q^{2n-\bar w-\wp}\right)^{-1}\right]
.
\end{split}
\end{equation}

Since we are interested in open string vertex operators that are inserted on the boundaries of the cylinder, we 
need to restrict the above propagator to the boundaries to get the relevant propagator. Evaluated at boundary 
points,  it is 
\begin{alignat}{2}
\pr(y, \yp )&=
\oh\ap g^{-1}\ln q-2\ap G^{-1}\ln\left[q^{\frac{1}{4}}
\vartheta_4\left(\frac{\vert y-\yp\vert}{2\tau },\frac{i}{\tau}\right)/
D(\tau)\right]
\quad& \text{for}\quad x\not=x^{\prime}
\quad,
\label{pr-n}\\
\pr(y, \yp )&=\pm\oh i\theta
\epsilon_{\zm}
(y-\yp)-2\ap G^{-1}
\ln\left[
\vartheta_1\left(\frac{\vert y-\yp\vert}{2\tau },\frac{i}{\tau}\right)/
D(\tau)\right]
\quad& \text{for}\quad  x=x^{\prime}
\quad,
\label{pr-p}
\end{alignat}
where $\pm$ correspond to $x=1$ and $x=0$, 
respectively. $\vartheta_1$ and $\vartheta_4$ are the Jacobi theta functions. We have also added 
constants to the propagators. They will be set to a convenient value in the next section. The $\epsilon_{\zm}$ 
function excludes the zero mode contribution. Explicitly, it is given by 
\begin{equation}\label{ep-o}
\epsilon_{\zm}(y)=\frac{i}{\pi}\ln\left(\frac{1-q^{-iy}}{1-q^{iy}}\right)=\epsilon(y)-\frac{y}{\tau}
\quad.
\end{equation}

In fact, what we have found above considerably differs from what we had in section 1 where the theory 
(its open string sector) is completely described in terms of the open string parameters. In our present 
discussion, we have obtained that the closed string metric $g$ no longer decouples. This fact has important 
consequences that we will discuss in the next section.

One basic question we should ask now is how to determine the modular measure $[d\tau]_B$ of open bosonic 
string with a constant $B$-field in arbitrary dimension. For this it is crucial that the measure 
factorises as $f(B)[d\tau]_0$, where the $B$-dependent factor is given by
\begin{equation*}
f(B)=\det(1+2\ai g^{-1}B)
\quad.
\end{equation*}
It was found by direct calculation in \cite{AC}. From Eq.\eqref{nv}we 
get $\sqrt{\det g}f(B)=\sqrt{\det G}$ (see also \cite{A}). Moreover, it was suggested in \cite{AD} that 
the original theory whose 
action is given by \eqref{ac0} or \eqref{ac} can be also described by a simpler action  
\begin{equation}\label{ac-1}
S=\frac{1}{4\ai }\int_{C_2}d^2z\,
G_{ij}\pd_aX^i\pd^aX^j
\end{equation} 
while correlation functions of the vertex operators include the build in star products ($\theta$-dependence). 
It is easy to see that it indeed works at the tree-level where the ansatz is simply 
\begin{equation}\label{an}
\langle V_1\ast V_2\ast\dots\ast V_N\rangle
\quad.
\end{equation}
 In fact, it follows because the corresponding Chan-Paton factor is $Tr(\lambda_1\dots
\lambda_N)$ \footnote{A motivation for this analogy is due to matrix (reduced) models where a map from 
a matrix to a function $A\rightarrow a(x)$ leads to $AB\rightarrow a(x)\ast b(x)$ (see, e.g., \cite{Kaw} and 
references therein).} \footnote{It is well-known that string theory imposes restrictions on possible 
gauge groups introduced via the Chan-Paton method (see, e.g., \cite{gsw} and 
references therein). In the problem at hand the $B$-field may lead to a potential clash with the 
Chan-Paton method. So the question arises whether such a method is still consistent. We will show in the 
appendix  that this is the case. However there exists the only allowed gauge group namely, $U(N)$. }. After 
this is understood, it becomes clear that the $\theta$-dependence is more 
involved on higher loop levels where there are products of traces. For example, in the case of interest the 
Chan-Paton factor is $Tr(\lambda_1\dots\lambda_N)Tr(\lambda_{N+1}\dots\lambda_M)$. For 
$N=0$ or $M-N=0$ that corresponds to planar diagrams, the ansatz reduces to the above one. 
This also follows 
from the corresponding propagator \eqref{pr-p}. It is clear that for $N\not=0,\,M-N\not =0$ that 
corresponds to non-planar diagrams, the ansatz \eqref{an} has to be modified. In this case the 
propagators \eqref{pr-n}-\eqref{pr-p} say us what to do. However, the measure is completely defined by the 
action \eqref{ac-1} \footnote{In a general case one also has to add a constant dilaton field as 
$\chi\hat\varphi$. Here $\chi$ is the Euler characteristic.}. 

Thus the problem reduced to the old one namely, how to extend the modular measure of open string to 
arbitrary dimension. This has been much studied to compute perturbative field 
theory amplitudes via string theory in the $\ap\rightarrow 0$ limit (see, e.g., \cite{DiV-rev}). In making our 
further analysis, we will adopt the proposal in \cite{Pobook} according to which the interpretation of each 
factor is transparent. The measure is thus 
\begin{equation}\label{me}
\int [d\tau]_0=\int_0^{\infty}\frac{d\tau}{\tau}\tau^{-\frac{d}{2}}
\left[\eta(i\tau)\right]^{2-d}
\quad,
\end{equation}
where $\eta$ is the Dedekind eta function.

The question that immediately arises is whether this measure could be obtained directly from a calculation, 
for instance, along the lines of \cite{Pobook}. At the tree-level, the commutation relations for the modes 
of $X^i$ that follow from the action \eqref{ac-1}
are as usual namely,  
\begin{equation}\label{com}
[\text{\footnotesize  X}^i, p^j]=2i\ap G^{ij}
\quad,\quad
[\alpha^i_n,\alpha^j_m]=2\ap n\delta_{n+m.0}G^{ij}
\quad,
\end{equation}
where $p^i\equiv\alpha^i_0$. Moreover, it turns out that the Virasoro generators don't depend on the zero 
modes {\footnotesize  X}. So, one can formally repeat the standard analysis to build the physical states via 
states in the Fock space. Moreover, it also means that the tachyonic vertex operator is given 
by \eqref{tach}. $d-2$ in \eqref{me} assumes that the reparametrization ghosts are included.


The scattering amplitudes can be defined in two ways. One can modify the ansatz \eqref{an} to do it 
consistent with the $\theta$ dependence that follows from the propagators or  it can be done by 
\begin{equation}\label{amp}
\begin{split}
A_{N.M}&=A(k_1,\dots ,k_N;k_{N+1},\dots ,k_M)\\
&={\cal N}_1\left(\ap\right)^{\Delta}
G_s^M\,Tr(\lambda_1\dots\lambda_N)\,Tr(\lambda_{N+1}\dots\lambda_M)\,{\cal V}_d
\langle V(k_1)\dots V(k_M)\rangle \\
&\phantom{=}+\text{noncyclic permutations}\quad,\quad
\langle \,\dots\,\rangle=\int [d\tau]_0\int {\cal D}X^{\prime}\,e^{-S}
\quad.
\end{split}
\end{equation}
Here $N$ vertex operators are attached to one boundary and $M-N$ vertex operators to the other one as in 
figure 1. ${\cal N}_1$ is a proper normalization constant (see \cite{DiV}). The correlation functions 
of exponential operators are simply computed by using the explicit form of the propagator. 

So, the amplitudes are 
\begin{equation}\label{amp-d}
\begin{split}
A_{N.M}&=
{\cal N}_1(2\pi )^d\left(\ap\right)^{\Delta}
G_s^M\,Tr(\lambda_1\dots\lambda_N)\,Tr(\lambda_{N+1}\dots\lambda_M)
\delta\left(\sum_{i=1}^Mk_i\right)
\int_0^{\infty}\frac{d\tau}{\tau}\tau^{-\frac{d}{2}}
\left[\eta(i\tau)\right]^{2-d}\, q^{\oh\ap\mathbf{k}g^{-1}\mathbf{k}}
\\
&\times
\prod_i^M\int_0^{y_{i-1}}dy_i  
\prod_{i=1}^N\prod_{j=N+1}^M
\left[q^{\frac{1}{4}}
\frac{\vartheta_4\left(\frac{\vert y_{ij}\vert}{2\tau},\frac{i}{\tau}\right) }
{D(\tau)}\right]^{2\ap k_iG^{-1}k_j}
\prod_{i<j}^N
\ep^{-\oh i\epsilon_{\zm}(y_{ij})k_i\theta k_j}
\left[
\frac{\vartheta_1\left(\frac{\vert y_{ij}\vert }{2\tau},\frac{i}{\tau}\right) }
{D(\tau)}\right]^{2\ap k_iG^{-1}k_j}
\\
&\times
\prod_{N+1\atop i<j}^M
\ep^{\oh i\epsilon_{\zm}(y_{ij})k_i\theta k_j}
\left[
\frac{\vartheta_1\left(\frac{\vert y_{ij}\vert}{2\tau},\frac{i}{\tau}\right) }
{D(\tau)}\right]^{2\ap k_iG^{-1}k_j}
\quad+\quad\text{noncyclic permutations}
\quad,
\end{split}
\end{equation}
where $y_{ij}=y_i-y_j $ and $\mathbf{k}=\sum_{i=1}^Nk_i$.

%
\vspace{-0.2cm}
\begin{figure}[ht]
\begin{center}
\includegraphics{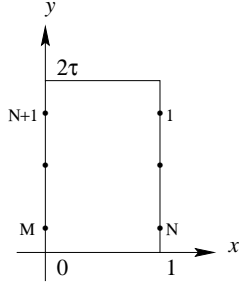}
\caption{The ordering of open string vertex operators on the cylinder.}
\label{fig:graph1}
\end{center}
\end{figure}
\vspace{-0.3cm} 

To complete the story, perhaps we should point out that the $\theta$-dependence of planar 
diagrams is very simple. It reduces to an overall phase factor 
$\ep^{-\oh i\sum_{i<j}^N k_i\theta k_j}$ i.e., the same as for planar diagrams of noncommutative 
field theories \cite{F}.


\subsection{Noncommutative field theory limit}

As discussed in the previous section, the noncommutative field theory limit is defined by 
$\ap\rightarrow 0$ at fixed $m, G$ and $\theta$ 
(i.e. $g^{-1}=-\frac{1}{(2\ai )^2}\theta G\theta$ via Eq.\eqref{nv}). On the other hand, it is well-known 
that only the neighbourhood of $q=1$ or, equivalently $\tau\rightarrow\infty$, contributes to the 
field theory limit (see, e.g., \cite{DiV-rev}). The restriction to the $\tau\rightarrow\infty$ edge of the 
moduli space remains true also in the noncommutative field theory limit since finite values of 
$\tau$ are even more suppressed due to the factor $q^{\oh \ap\mathbf{k}g^{-1}\mathbf{k}}$ in 
Eq.\eqref{amp-d}. 


The noncommutative field theory limit is also simply carried out by sending both $\tau$ and $y$ to infinity 
while keeping variables \footnote{We disregard the pinching configurations that result in the one-particle 
reducible diagrams.}
\begin{equation}\label{pt}
t=2\ai\tau
\quad,\quad
\nu_i=y_i/2\tau
\end{equation}
finite.  
 
In this limit, the propagators become \footnote{We have set 
$D(\tau)=\tau^{-1}\left[\eta(i/\tau)\right]^3$.}
\begin{alignat}{2}
\pr(\nu, \pnu )&=
\frac{1}{4t}\theta G\theta
+G^{-1}t\left[\left(\nu-\pnu\right)^2-\left\vert\nu-\pnu\right\vert\right]
\quad& \text{for}\quad x\not=x^{\prime}
\quad,
\label{ft-n}\\
\pr(\nu, \pnu )&=\pm\oh i\theta\epsilon_{\zm}(\nu-\pnu)
+G^{-1}t\left[\left(\nu-\pnu\right)^2-\left\vert\nu-\pnu\right\vert\right]
\quad& \text{for}\quad  x=x^{\prime}
\quad.
\label{ft-p}
\end{alignat}
At this point a couple of comments are in order.
\newline (i) The $\epsilon_{\zm}$ function defined by Eq.\eqref{ep-o} is related via 
$\frac{\pd }{\pd y}\epsilon_{\zm} (y)=2\delta_{\zm}(y)$ to the delta function that excludes the 
zero mode i.e., $\delta_{\zm}(y)=\delta(y)-1/2\tau$. Naively, 
the difference between these $\epsilon$ functions disappears in the $\tau\rightarrow\infty$ limit. In fact, 
this is not the case because the last term survives as far as the rescaling \eqref{pt} is taken into account. 
Thus 
\begin{equation}\label{ep}
\epsilon_{\zm}(\nu)=\epsilon(\nu)-2\nu
\quad.
\end{equation}
\noindent (ii) The difference between $\epsilon$ and $ \epsilon_{\zm}$ indeed disappears for all 
planar amplitudes as well as tree level amplitudes. In this case the zero mode contribution to the 
amplitudes $\sum_{i<j}p_i\theta p_j \nu_{ij}$ vanishes due to the momentum conservation $\sum p_i=0$. 

Thus, in the field theory limit the amplitudes \eqref{amp-d} become
\begin{equation}\label{amp-ft}
\begin{split}
A_{N.M}&=
{\cal N}_1^{\prime}\,\g^M\,Tr(\lambda_1.\,.\,.\lambda_N)\,Tr(\lambda_{N+1}.\,.\,.\lambda_M)
\delta\left(\sum_{i=1}^Mk_i\right)
\prod^N_{i<j}\ep^{-\frac{i}{2}k_i\theta k_j}
\prod_{N+1\atop i<j}^M\ep^{\frac{i}{2}k_i\theta k_j}
\int_0^{\infty}\frac{dt}{t}\,t^{M-\frac{d}{2}}
\,\ep^{-m^2t-\mathbf{k}\circ\mathbf{k}/t}
\\
&\times
\prod_i^M\int_0^{\nu_{i-1}}d\nu_i  
\prod_{i<j}^M \ep^{t(\left\vert\nu_{ij}\right\vert-\nu^2_{ij})k_iG^{-1}k_j}
\prod_{i<j}^N\ep^{i\nu_{ij}k_i\theta k_j}
\prod_{N+1\atop i<j}^M\ep^{-i\nu_{ij}k_i\theta k_j}\quad +\quad\text{noncyclic permutations}
\quad,
\end{split}
\end{equation}
where $\mathbf{k}\circ\mathbf{k}=-\frac{1}{4}\mathbf{k}\theta G\theta\mathbf{k}$. To get 
this form we introduced the mass for the open string 
tachyon as $m^2=(2-d)/24\ap$ and used the relation between the open string coupling constant and 
the $\3$ theory coupling constant. 

We will conclude this subsection with a couple of examples to illustrate the use of the string amplitudes in 
practical calculations of one loop Feynman diagrams of the noncommutative $\3$ theory.

{\it Example 1}. As a warmup, let us consider the simplest nonplanar amplitude and, as by-product, 
reproduce a result of \cite{MRS}. The amplitude (modulo a normalization constant) is given by
\begin{equation}\label{12}
A_{1.2}=\g^2\,Tr(\lambda_1)\,Tr(\lambda_2)
\delta\left(k_1+k_2\right)
\int_0^{\infty}dt\, t^{1-\frac{d}{2}}\,
\ep^{-m^2t-k_1\circ k_1/t}
\int_0^1d\nu_1\, 
\ep^{t(\nu_1^2-\nu_1)k_1G^{-1}k_1}\,
\quad.
\end{equation}
Here we fix the translational invariance on the cylinder by setting the second vertex operator at the origin i.e., 
$\nu_2=0$. 

The next step in finding the correspondence with the field theory diagram is to introduce new integration 
variables which correspond to the Schwinger parameters
\begin{equation}\label{sch}
\alpha_1=t\nu_1
\quad,\quad
\alpha_2=t(1-\nu_1)
\quad.
\end{equation}
In terms of these variables, the amplitude is written as 
\begin{equation}\label{12-ft}
A_{1.2}=\g^2\,Tr(\lambda_1)\,Tr(\lambda_2)
 \delta\left(k_1+k_2\right)
\int\limits_0^{\infty}\int\limits_0^{\infty}
\frac{d\alpha_1d\alpha_2}{\left(\alpha_1+\alpha_2\right)^{\frac{d}{2}}}\,
\ep^{-m^2\left(\alpha_1+\alpha_2\right)}\,
\ep^{-\frac{k_1\circ k_1 }{\left(\alpha_1+\alpha_2\right)}}\,
\ep^{-\frac{\alpha_1\alpha_2}{\left(\alpha_1+\alpha_2\right)}k_1G^{-1}k_1}
\quad.
\end{equation}
  
It is now clear that what we have found is exactly the simplest nonplanar Feynman diagram of the 
noncommutative $\3$ theory as shown in Fig.2.

\newpage
%
\vspace{-0.2cm}
\begin{figure}[ht]
\begin{center}
\includegraphics{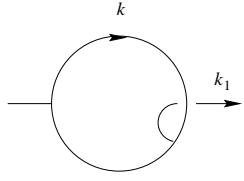}
\caption{The simplest nonplanar diagram of the noncommutative $\3$ theory.}
\label{fig:graph2}
\end{center}
\end{figure}
\vspace{-0.3cm} 

Before we give the next example, we first want to make one important remark. The point is that the 
second exponent, in fact, acts as an effective regulator of the otherwise UV divergent diagram. This 
interesting observation was made in \cite{CR, MRS} by original analysis of perturbation expansions 
of noncommutative theories. In \cite{MRS} it was also suggested that the effect is due to the closed string 
metric and a stringy interpretation was given. It is based on the known fact that for $B=0$ the UV 
limit ($\tau\rightarrow 0$) 
of the open string channel corresponds to the IR limit for the closed string channel. What we found starting 
directly from open string theory with the $B$-field is a little bit different. Indeed, the effect is due to 
the closed string metric but it has nothing common with the $\tau\rightarrow 0$ limit for open string. 
As we have seen the crucial point is that the closed string metric $g$ does not decouple from the propagator 
between boundary points in the noncommutative field theory limit i.e., $\tau\rightarrow\infty$! So, 
in the noncommutative field theory limit there is effectively a signal of closed string sector.
  

{\it Example 2}. The next example we would like to consider is the $A_{2.4}$ string amplitude. Here we fix 
the translational invariance on the cylinder by setting the fourth vertex operator at the origin i.e., 
$\nu_4=0$. We now can replace the multiple integral by a sum of ordered multiple integrals. This is 
clear just by substituting the expansion of unity that for the problem at 
hand (for ordered $\nu_1$ and $\nu_2$) is 
$1=H(\nu_{23})+H(\nu_{13})H(\nu_{32})+H(\nu_{31})$ \footnote{$H$ means the 
Heaviside step function.}. Now we can proceed along the lines of \cite{DiV-rev} i.e., we rewrite the 
ordered integrals via the standard integrals over the Schwinger parameters. Let us explicitly illustrate 
how it works for the first term where the ordering is $\nu_1\geq\nu_2\geq\nu_3$. In this case 
 the corresponding contribution (modulo a normalization constant) is given by 
\begin{equation}\label{24}
\begin{split}
&\phantom{.}
\g^4\,Tr(\lambda_1\lambda_2)\,Tr(\lambda_3\lambda_4)
\delta\left(\sum_{i=1}^4k_i\right)
\,\ep^{-\frac{i}{2}k_1\theta k_2+\frac{i}{2}k_3\theta k_4}
\int_0^{\infty}dt\, t^{3-\frac{d}{2}}\,\ep^{-m^2t-\mathbf{k}\circ\mathbf{k}/t}
\\
&\times
\int_0^1d\nu_1\int_0^{\nu_1}d\nu_2 \int_0^{\nu_2}d\nu_3
\,\ep^{i\nu_{12}k_1\theta k_2-i\nu_{3}k_3\theta k_4}
\prod_{i=1}^3\ep^{t(\nu_i^2-\nu_i)k_iG^{-1}k_i}
\prod_{i<j}^3\ep^{2t(\nu_i\nu_j-\nu_j)k_iG^{-1}k_j}
\quad.
\end{split}
\end{equation}
The Schwinger parameters can be introduced as 
\begin{equation}\label{sch-2}
\alpha_1=t\nu_{12}\quad,\quad
\alpha_2=t\nu_{23}\quad,\quad
\alpha_3=t\nu_3\quad,\quad
\alpha_4=t(1-\nu_1)
\quad.
\end{equation}
So, the expression \eqref{24} becomes
\begin{equation}\label{24-1234}
\begin{split}
&\phantom{.}\g^4\,Tr(\lambda_1\lambda_2)\,Tr(\lambda_3\lambda_4)
\delta\left(\sum_{i=1}^4k_i\right)
\,\ep^{-\frac{i}{2}k_1\theta k_2+\frac{i}{2}k_3\theta k_4}
\prod_{i=1}^4\int_0^{\infty}d\alpha_i\, \alpha^{-\frac{d}{2}}\,\ep^{-m^2\alpha}
\,\ep^{-\frac{1}{\alpha}\mathbf{k}\circ\mathbf{k}}
\,\ep^{\frac{i}{\alpha}\left(\alpha_1k_1\theta k_2-\alpha_3k_3\theta k_4\right)}
\\
&\times
\ep^{\frac{1}{\alpha}
\left(-\alpha_1\alpha_4k_1G^{-1}k_1-\alpha_1\alpha_2k_2G^{-1}k_2-
\alpha_2\alpha_3k_3G^{-1}k_3-\alpha_3\alpha_4k_4G^{-1}k_4-
\alpha_1\alpha_3(k_2+k_3)G^{-1}(k_2+k_3)-\alpha_2\alpha_4kG^{-1}k
\right)}
\quad,
\end{split}
\end{equation}
where $\alpha=\sum_{i=1}^4\alpha_i $. It is straightforward to get the Schwinger 
representation for the other terms. As a result, the amplitude $A_{2.4}$ reduces to a sum of three 
nonplanar Feynman diagrams of the noncommutative $\3$ theory as shown in fig.3.

%
\vspace{0.2cm}
\begin{figure}[ht]
\begin{center}
\includegraphics{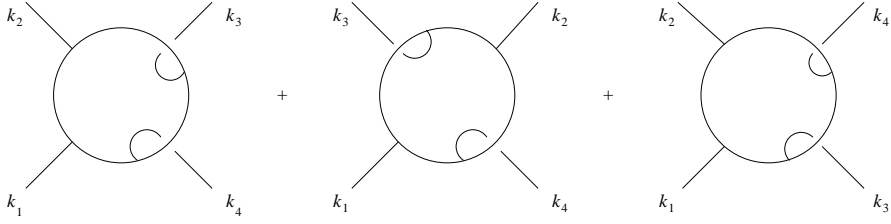}
\caption{Three nonplanar field theory diagrams that correspond 
to the $A_{2.4}$ string amplitude.}
\label{fig:graph3}
\end{center}
\end{figure}
\vspace{-0.3cm} 

\subsection{$\tau\rightarrow 0$ limit}

Now we consider another limit: $\tau\rightarrow 0$ while keeping all other parameters fixed. It is 
well-known that for $B=0$ this is the open string UV limit. Moreover, it is interpreted as a long-distance 
effect because the leading asymptotics are given by the closed string tachyon as well as the lightest closed 
string states. So, our purpose is to analyze what happens in the presence of the $B$-field.

In this limit, the propagators \eqref{pr-n}-\eqref{pr-p} become 
\begin{alignat}{2}
\pr(y, \yp )&=
-\frac{\ai}{2\tau} g^{-1}-2\ap G^{-1}\ln\tau
\quad& \text{for}\quad x\not=x^{\prime}
\quad,
\label{pr-n0}\\
\pr(y, \yp )&=\pm\oh i\theta
\epsilon_{\zm}(y-\yp)+
2\ap G^{-1}P(y-\yp )-
2\ap G^{-1}\ln\tau
\quad& \text{for}\quad  x=x^{\prime}
\quad.
\label{pr-p0}
\end{alignat}
Here $P(y)=\sum_{n=1}^{\infty}\frac{1}{n}\cos \frac{\pi n}{\tau}y$. As in the $\tau\rightarrow\infty$ 
case that we discussed first, it is useful to define new variables 
\begin{equation}\label{pt-0}
t=\frac{\ai}{2\tau}
\quad,\quad
\nu_i=y_i/2\tau
\quad.
\end{equation}
Then we get the amplitudes (modulo a normalization constant)
\begin{equation}\label{amp-d0}
\begin{split}
A_{N.M}&=
\left(\ap\right)^{\Delta-1}
\delta\left(\sum_{i=1}^Mk_i\right)
\left[
G_s^N\,Tr(\lambda_1.\,.\,.\lambda_N)\prod_{i=1}^{N-1}\int_0^{\nu_{i-1}}d\nu_i
\prod_{i<j}^N\ep^{-\oh i\epsilon_{\zm}(\nu_{ij})k_i\theta k_j}
\, 
\ep^{-2\ap k_iG^{-1}k_jP(\nu_{ij})}
\right]_{\nu_N=0} \\
&\times
\frac{1}{\mathbf{k}g^{-1}\mathbf{k}+\mathbf{m}^2}
\left[
G_s^{M-N}\,Tr(\lambda_{N+1}.\,.\,.\lambda_M)\prod_{i=N+1}^{M-1}\int_0^{\nu_{i-1}}d\nu_i
\prod_{N+1\atop i<j}^M\ep^{\oh i\epsilon_{\zm}(\nu_{ij})k_i\theta k_j}
\ep^{-2\ap k_iG^{-1}k_jP(\nu_{ij})}
\right]_{\nu_M=0} \\
&+\quad\text{noncyclic permutations}
\quad.
\end{split}
\end{equation}
The translational invariance is now fixed by setting $\nu_N=\nu_M=0$. We also introduce the mass for 
the closed string tachyon as $\mathbf{m}^2=(2-d)/6\ap$. 

What we see from the above is that in the $\tau\rightarrow 0$ limit for $B\not=0$  the amplitudes factorize 
as in the case $B=0$. There is no new effect here. So, the interpretation is the standard one as a long-distance 
effect with the asymptotics due to the closed string states.

\section{Concluding Comments } 
\renewcommand{\theequation}{3.\arabic{equation}}
\setcounter{equation}{0} 
What we have learned is that in the noncommutative field theory limit of open string theory with 
the $B$-field at the one loop level there is a signal of closed string sector. To be more precise, 
we found that $\theta$ appears not only via the $\ast$-product as it does at the tree level but 
via $-\frac{1}{4}\theta G\theta$ that corresponds to the closed string metric $g$. A recent analysis 
of noncommutative field theories 
assumes that some closed string modes already appear there \cite{RS}. We do not see this explicitly 
in our analysis of the noncommutative field theory limit where we found only the closed string 
parameters rather than the modes. However, to be cautious, we should mention that we did not discuss 
singularities and their regularization. 

It is important to emphasize that the factor $q^{\oh\ap\mathbf{k}g^{-1}\mathbf{k}}$, or equivalently 
the $g$-dependence of amplitudes, which is crucial for 
the noncommutative field theory limit of nonplanar diagrams is universal. It does not depend on the kind of 
vertex operators. This is clear because the effect is due to the first term in the propagator \eqref{pr-n}. 
It is a constant, so contributions come only from exponents.

From the physical point of view it is more interesting to consider the noncommutative gauge 
theory. In fact, at the one-loop level it can be done along the lines of the present paper by considering 
the vertex operator for a gauge field 
\begin{equation}\label{vec}
V(\xi, k)=\int ds\, \xi\cdot \pd_ sX\,\ep^{ik\cdot X}
\end{equation}
instead of the tachyon operator \eqref{tach}. Then, the corresponding amplitudes are computed 
by using the propagators \eqref{pr-n}-\eqref{pr-p} within the point splitting renormalization 
scheme \cite{SW}. We hope to return to this important problem in the near future \cite{AD1}.

\vspace{.25cm} {\bf Acknowledgments}

\vspace{.25cm} 
We would like to thank A.A. Tseytlin for a collaboration at an initial stage, and also for useful 
comments and reading the manuscript. The work of O.A. is supported in part by the Alexander 
von Humboldt Foundation and by Russian Basic Research Foundation under Grant No. 9901-01169.
The work of H.D. is supported in part by DFG.


\vspace{.25cm} {\bf Appendix}
\renewcommand{\theequation}{A.\arabic{equation}}
\setcounter{equation}{0}

\vspace{.25cm} 
The purpose of this appendix is to show how restrictions on gauge groups arise within open string theory in 
the presence of the $B$-field. In general, there are a variety of ways to do this. Here we will focus on the 
classical recipe \cite{gsw} which is based on simple factorization properties of string amplitudes. So, let us take 
the tree amplitude \eqref{amp0} and look at its factorization $A_M\rightarrow A_P\,A_{M-P}$. It is 
well-known that the only novelty due to the $B$-field is a phase factor 
\begin{equation}\label{phase}
{\cal P}_1^M=\prod_{\substack{1\\ i<j}}^M\ep^{-\frac{i}{2}k_i\theta k_j}
\end{equation}
that appears in the expression for the amplitude. The crucial fact for what follows is that ${\cal P}_1^M$ due 
to momentum conservation obeys the factorization relation 
\begin{equation}\label{fac}
{\cal P}_1^M={\cal P}_1^P\,{\cal P}_{P+1}^M
\quad.
\end{equation}
With above formula the generalization of the classical analysis is straightforward. As a result, Eq. (6.1.11) 
of \cite{gsw} becomes 
\begin{equation}\label{fac2}
\begin{split}
Tr\Biggl[\Bigl(\lambda_1\dots&\lambda_P\,{\cal P}_1^P
-
(-)^P\lambda_P\dots\lambda_1\left({\cal P}_1^P\right)^{-1}\Bigr)
\Bigl(\lambda_{P+1}\dots\lambda_M\,{\cal P}_{P+1}^M
-
(-)^{M-P}\lambda_M\dots\lambda_{P+1}\left({\cal P}_{P+1}^M\right)^{-1}\Bigr)\Biggr] \\
=&\sum_{\alpha}Tr\Biggl[\Bigl(\lambda_1\dots\lambda_P\,{\cal P}_1^P
-
(-)^P\lambda_P\dots\lambda_1\left({\cal P}_1^P\right)^{-1}\Bigr)\lambda_{\alpha}\Biggr] \\
&\times Tr\Biggl[\lambda_{\alpha}^{\text{\tiny T}}
\Bigl(\lambda_{P+1}\dots\lambda_M\,{\cal P}_{P+1}^M
-
(-)^{M-P}\lambda_M\dots\lambda_{P+1}\left({\cal P}_{P+1}^M\right)^{-1}\Bigr)\Biggr] 
\quad.
\end{split}
\end{equation}
This equation is satisfied if a matrix 
\begin{equation}\label{sol}
\lambda=\lambda_1\dots
\lambda_P\,{\cal P}_1^P
-(-)^P\lambda_P\dots\lambda_1\left({\cal P}_1^P\right)^{-1}
\end{equation}
belongs to the algebra of the matrices $\lambda_i$. It is known that in the case of $\theta=0$ the allowed 
gauge groups are $U(n),\,SO(n)$ and $USp(2n)$. To see what survives for nonzero $\theta$ let us specialize 
to  $P=2$. Then the following direct algebra
\begin{alignat}{2}
\lambda^{\dagger}&=-\left(\lambda_1\lambda_2\,\ep^{-\frac{i}{2}k_1\theta k_2}-
\lambda_2\lambda_1\,\ep^{\frac{i}{2}k_1\theta k_2}\right)=-\lambda &
\quad&\text{for}\quad \lambda_i\in u(n)\quad,\\
\lambda^{\text{\tiny T}}&=-\left(\lambda_1\lambda_2\,\ep^{\frac{i}{2}k_1\theta k_2}-
\lambda_2\lambda_1\,\ep^{-\frac{i}{2}k_1\theta k_2}\right)\not=-\lambda &
\quad&\text{for}\quad \lambda_i\in so(n)\quad, \\
\lambda^{\text{\tiny T}}&=s^{-1}\left(\lambda_2\lambda_1\,\ep^{-\frac{i}{2}k_1\theta k_2}-
\lambda_1\lambda_2\,\ep^{\frac{i}{2}k_1\theta k_2}\right)s\not=-s^{-1}\lambda s&
\quad&\text{for}\quad \lambda_i\in usp(n),\,\,\text{where}\,\,
\lambda_i^{\text{\tiny T}}=-s^{-1}\lambda_i s\quad
\end{alignat}
shows that the only surviver is $U(n)$. The latter is in harmony with the result obtained within 
noncommutative gauge theory claiming that the noncommutative gauge transformation is consistent only 
for the unitary group $U(n)$ (see e.g., \cite{Wess}).



\end{document}